\documentclass[12pt]{article}
\usepackage{amsmath,amssymb,epsfig, bm,color,psfrag,pstool,subcaption}
\usepackage{feynmf,graphicx}

\allowdisplaybreaks

\newcommand{\order}{{\cal O}}

\newcommand{\be}{\begin{equation}}  
\newcommand{\ee}{\end{equation}} 

\makeatletter
\newcommand{\vast}{\bBigg@{4}}
\newcommand{\Vast}{\bBigg@{5}}
\makeatother

\begin{document}
  
\begin{titlepage}

\begin{flushright}
  FERMILAB-PUB-17-501-T \\
  WSU-HEP-1713 \\
  November 2, 2017
\end{flushright}

\vspace{0.7cm}
\begin{center}
\Large\bf 
Reply to Comment [arXiv:1708.09341] by Birse and McGovern on
``Nucleon spin-averaged forward virtual Compton tensor at large $Q^2$''
\end{center}

\vspace{0.8cm}
\begin{center}
{\sc  Richard J. Hill$^{(a)}$ and Gil Paz$^{(b)}$  } \\
\vspace{0.4cm}
{\it 
  $^{(a)}$
  Department of Physics and Astronomy, University of Kentucky, Lexington, KY 40506, USA, and \\
  Fermilab, Batavia, IL 60510, USA
}\\
\vspace{0.7cm}
{\it 
$^{(b)}$ 
Department of Physics and Astronomy \\
Wayne State University, Detroit, Michigan 48201, USA}
\end{center}
\vspace{1.0cm}

\begin{abstract}
\vspace{0.2cm}
\noindent  

We reply to misleading claims made in a comment to our work by
Birse and McGovern in arXiv:1708.09341.

\vfil
\end{abstract}
\end{titlepage}

\section{Review}

In Ref.~\cite{Hill:2016bjv} we recently computed, for the first time,
the leading $1/Q^2$ behavior at large $Q^2$ of the quantity $W_1(0,Q^2)$ appearing in
the spin-averaged forward virtual Compton tensor.  This function plays an
important role in the two-photon exchange contribution to the muonic
hydrogen Lamb shift.  This contribution provides the dominant theoretical
uncertainty in the determination of the proton charge radius, and of the
Rydberg constant, one or our most precisely known fundamental constants.

The OPE evaluation of $W_1(0,Q^2)$ has a long history, and our work
corrects the decades old result of Collins~\cite{Collins:1978hi}.  In
that work, an incorrect quark charge factor led to an overestimate of the
spin-0 contribution by an order of magnitude.  In reality, the leading
OPE evaluation of $W_1(0,Q^2)$ is dominated by the spin-2 contribution, which
was not considered in Ref.~\cite{Collins:1978hi}.

With a correct determination of $W_1(0,Q^2)$ at large $Q^2$ in hand,
we ended our paper with a discussion of the implications for the muonic
hydrogen Lamb shift.  As summarized by Fig.~9 in Ref.~\cite{Hill:2016bjv},
we presented two results: the first, based on Fig.~8 in Ref.~\cite{Hill:2016bjv},
was an interpolation using the OPE evaluation at
large $Q^2$ and only the constraints~\cite{Hill:2011wy} of NRQED at
low $Q^2$.
The second result, based on Fig.~11 in Ref.~\cite{Hill:2016bjv},
was an interpolation using the same OPE evaluation at
large $Q^2$, and including chiral lagrangian constraints
(in fact, the evaluation of Birse and McGovern in Ref.~\cite{Birse:2012eb})
to extend the low $Q^2$ range.

\section{Remarks}

In an arXiv posting, arXiv:1708.09341, Birse and McGovern (BM) make several misleading
claims related to our work.  We reply with the following remarks. 

\vspace{5mm}
(i)
In a preface to their discussion, BM claim that the OPE computation~\cite{Hill:2016bjv}
validates their computation~\cite{Birse:2012eb}.
It is difficult to interpret this statement. 
In fact, the central value for the coefficient of $1/Q^2$ that was deduced in
Ref.~\cite{Birse:2012eb} differs by a large factor, $\sim 3 - 4$, from the correct
result, as BM themselves admit.
The disagreement can be mitigated by accounting for uncertainties
in the treatment of higher order chiral corrections; surprisingly, the authors
later criticize precisely this method for estimating uncertainties in the
evaluation of the Lamb shift (see point (iv) below). 
Regardless of the numerical comparison, the claimed relation between the
low-energy quantity proportional to $M_\beta^4 \beta$ of Ref.~\cite{Birse:2012eb}, 
and the perturbative OPE expression,
given by quark and gluon matrix elements, 
has not been justified to be a valid QCD relation.

\vspace{5mm}
(ii)
The bulk of the analysis in the comment by BM concerns
a spurious ``apples to oranges'' comparison of our determination in
Fig.~9 that used only
NRQED without chiral lagrangian constraints, to their determination
that used chiral lagrangan constraints.
If they had instead employed an ``apples to apples'' comparison of
our determination in Fig.~9 that used chiral lagrangian constraints from Fig.~11,
then agreement is obtained at a level consistent with remaining analysis differences.
This can be seen from Fig.~9 itself comparing
``[3]'' (Ref.~\cite{Birse:2012eb}) and ``Fig.~11''.   

\vspace{5mm}
(iii)
In their comment, BM also discuss an alternative utilization of
our OPE result at large $Q^2$, that would use only
the NRQED expansion at small $Q^2$.
By first subtracting a combination of elastic form
factors whose Taylor expansion cancels the dominant part of the
$\order(Q^2)$ contributions to $W_1(0,Q^2)$, the resulting interpolation
is argued to have smaller uncertainty.
We point out however that the uncertainties arising from moments of the elastic
form factors, such as $r_M$, are not avoided by
this procedure, but rather shuffled into a different part of
the calculation.

Let us remark on the subtraction of a term involving elastic form factors
from the total $W_1(0,Q^2)$.  When constrained by experimental data, the
Lamb shift contribution from any such term has
a controlled uncertainty, owing to the existence
of elastic form factor data throughout the entire $Q^2$ range of interest.
For example, while the magnetic radius may induce a large uncertainty
in the small-$Q^2$ Taylor expansion of $W_1(0,Q^2)^{\rm SIFF}$, this uncertainty remains
controlled when applied to the interpolation for the Lamb shift
involving complete form factors.

However,  there is not a unique
combination of elastic form factors whose Taylor expansion cancels
a chosen part of the $\order(Q^2)$ expansion of $W_1(0,Q^2)$. 
This ambiguity manifests itself in the ``Born subtraction''%
\footnote{
  We remark that the word ``Born'' used in this context in the
  literature refers to a conventional subtraction. 
  It does not have the usual technical meaning, i.e., as the
  leading expression for a physical quantity obtained by expressing
  the Hamiltonian in terms of an unperturbed problem plus a perturbation.
  The label ``proton pole'' or ``elastic'' used in the literature
  is also conventional: the very need for a subtraction implies that
  ${\rm Im}W_1(\nu,Q^2)$ does not uniquely determine a contribution
  from proton intermediate states in the dispersion relation for
  $W_1(\nu,Q^2)$. 
  The most common convention coincides with a form factor insertion
  ansatz, for which we use the more
  descriptive Sticking In Form Factors (SIFF)
  label.
}
performed by BM 
in Ref.~\cite{Birse:2012eb}, using third order expressions for
the elastic form factors from Ref.~\cite{Bernard:1995dp}.
At any given order in the chiral expansion (e.g. third order), 
different Born subtractions with the above-mentioned properties
may be constructed, which differ only by higher order terms than have
been calculated.  Choosing one amongst these possibilities amounts
to an implicit scheme or model dependence.
Phrased equivalently, the apparent
uncertainty reduction in $W_1(0,Q^2) - W_1(0,Q^2)^{\rm Born}$
is accompanied by an enlarged uncertainty in 
$W_1(0,Q^2)^{\rm Born}$. 

\vspace{5mm}
(iv)
Beyond the issue of Born subtraction, 
uncertainties from inputs and formalism should be accounted for in
the computation of $W_1(0,Q^2) - W_1(0,Q^2)^{\rm SIFF}$ using chiral lagrangian
analysis.
We did not attempt a formal quantification of these uncertainties
in the appendix of Ref.~\cite{Hill:2016bjv},
but simply displayed
a comparison between different orders in chiral power counting.  Although BM
appear to take issue with this identification of uncertainty, taking the difference
between successive orders in any power counting scheme is a standard method for
error estimation.

\end{document}